\documentclass[conference, 10pt]{IEEEtran}

\usepackage{amsmath,amsfonts}
\usepackage{algorithmic}
\usepackage{algorithm}
\usepackage{array}
\usepackage{textcomp}
\usepackage{url}
\usepackage{verbatim}
\usepackage{graphicx}
\usepackage{cite}
\usepackage{subcaption}
\usepackage{xcolor}

\usepackage{mathtools} 

\makeatletter

\newcounter{author}
\renewcommand{\author}[2][]{
   \stepcounter{author}
   \@namedef{author@\theauthor}{#2}
   \@namedef{authorlabel@\theauthor}{#1}
}

\newcounter{address}
\newcommand{\address}[2][]{
   \stepcounter{address}
   \@namedef{address@\theaddress}{#2}
   \@namedef{addresslabel@\theaddress}{#1}
}

\newcommand{\alsep}{and}

\def\newmaketitle{\par%
  \begingroup%
  \normalfont%
  \def\thefootnote{}
  \def\footnotemark{}
  \let\@makefnmark\relax
  \footnotesize
  \footnotesep 0.7\baselineskip
  \normalsize%
  \twocolumn[\thenewmaketitle\@IEEEaftertitletext]%
  \if@IEEEusingpubid
     \enlargethispage{-\@IEEEpubidpullup}%
  \fi
  \endgroup
  \setcounter{footnote}{0}\let\maketitle\relax\let\@maketitle\relax
  \gdef\@thanks{}%
  \let\thanks\relax
  }

\def\thenewmaketitle{
  \newpage
  \begin{center}%
    \vskip0.2em{\Huge\@IEEEcompsoconly{\sffamily}\@IEEEcompsocconfonly{\normalfont\normalsize\vskip 2\@IEEEnormalsizeunitybaselineskip
   \bfseries\large}\@title\par}\vskip1.0em\par%
    \vspace{1ex}
    \newcounter{c@author}
    \newcounter{c@tmp}
    \ifthenelse{\value{author}=2}{%
      \newcommand{\liand}{ and }}{%
      \newcommand{\liand}{, and }}
    \ifthenelse{\value{address}<2}{%
      \@nameuse{author@1}%
      \stepcounter{c@author}%
      \whiledo{\value{c@author}<\value{author}}{%
        \setcounter{c@tmp}{\value{author}}%
        \addtocounter{c@tmp}{-\value{c@author}}%
        \ifthenelse{\value{c@tmp}=1}{%
          \renewcommand{\alsep}{\liand}}{\renewcommand{\alsep}{, }}%
        \stepcounter{c@author}\alsep \@nameuse{author@\thec@author}}\\%
    }
    {
      \@nameuse{author@1}${}^{(\ref{\@nameuse{authorlabel@1}})}$%
      \stepcounter{c@author}%
      \whiledo{\value{c@author}<\value{author}}{%
      \setcounter{c@tmp}{\value{author}}%
      \addtocounter{c@tmp}{-\value{c@author}}%
      \ifthenelse{\value{c@tmp}=1}{%
        \renewcommand{\alsep}{\liand}}{\renewcommand{\alsep}{, }}%
      \stepcounter{c@author}\alsep \@nameuse{author@\thec@author}%
        ${}^{(\ref{\@nameuse{authorlabel@\thec@author}})}$%
      }
    }
    \vspace{0.2ex}

    \ifthenelse{\value{address}>0}{%
      \ifthenelse{\value{address}=1}{
        {\@nameuse{address@1}}
      }
      {
        \newcounter{c@address}

        \begin{center}
        \whiledo{\value{c@address}<\value{address}}
        {
          \refstepcounter{c@address}
            ${}^{(\thec@address)}$\,%
              \label{\@nameuse{addresslabel@\thec@address}}%
              \@nameuse{address@\thec@address}\\ %
        }
        \end{center}
      } 
    }
    {
      \relax
    }
  \end{center}
}

\makeatother

\title{GS-SBL: Bridging Greedy Pursuit and Sparse Bayesian Learning for Efficient 3D Wireless Channel Modeling}

\author[org1]{Mushfiqur Rahman}
\author[org1]{\.{I}smail G\"{u}ven\c{c}}
\author[org2]{David Matolak}

\address[org1]{Department of Electrical and Computer Engineering, North Carolina State University, Raleigh, NC 27606, USA}
\address[org2]{Department of Electrical Engineering, University of South Carolina, Columbia, SC 29208, USA}

\usepackage{tikz}
\newcommand\copyrighttext{%
  \footnotesize 979-8-3315-xxxx-x/26/\$31.00~\copyright~2026 IEEE. Personal use permitted. To appear in: 2026 IEEE AP-S/URSI.}
\newcommand\copyrightnotice{%
\begin{tikzpicture}[remember picture,overlay]
\node[anchor=south, yshift=15pt] at (current page.south) {\copyrighttext};
\end{tikzpicture}%
}

\begin{document}

\newmaketitle
\copyrightnotice

\begin{abstract}
Robust cognitive radio development requires accurate 3D path loss models. Traditional empirical models often lack environment-awareness, while deep learning approaches are frequently constrained by the scarcity of large-scale training datasets. This work leverages the inherent sparsity of wireless propagation to model scenario-specific channels by identifying a discrete set of virtual signal sources. We propose a novel Greedy Sequential Sparse Bayesian Learning (GS-SBL) framework that bridges the gap between the computational efficiency of Orthogonal Matching Pursuit (OMP) and the robust uncertainty quantification of SBL. 
Unlike standard top-down SBL, which updates all source hyperparameters simultaneously, our approach employs a ``Micro-SBL'' architecture. We sequentially evaluate candidate source locations in isolation by executing localized, low-iteration SBL loops and selecting the source that minimizes the $L_2$ residual error. Once identified, the source and its corresponding power are added to the support set, and the process repeats on the signal residual to identify subsequent sources. Experimental results on real-world 3D propagation data demonstrate that the GS-SBL framework significantly outperforms OMP in terms of generalization. By utilizing SBL as a sequential source identifier rather than a global optimizer, the proposed method preserves Bayesian high-resolution accuracy while achieving the execution speeds necessary for real-time 3D path loss characterization.
\end{abstract}


\section{Introduction}
Wireless node density in 3D space is surging, particularly with the widespread adoption of unmanned aerial vehicles~(UAVs)~\cite{wang2023sparse}. Operating in shared 3D environments, these heterogeneous devices require reliable, seamless connectivity. To mitigate the resulting spectrum scarcity, dynamic spectrum sharing via cognitive radio (CR) has emerged as a pivotal solution. Central to CR is the generation of an accurate radio environment map (REM), which utilizes field measurements and path loss models to provide a comprehensive spatial representation of power spectral density within a region of interest~\cite{shen20213d}. Existing path loss models generally fall into two categories: generic statistical models, such as the Okumura-Hata model, and environment-specific data-driven models, including deep learning and spatial interpolation techniques like Kriging or Inverse Distance Weighting (IDW)~\cite{wang2024sparse}. While statistical models often fail to account for complex 3D obstacles and site-specific building geometries, data-driven approaches suffer from significant computational overhead or prohibitive storage requirements. In this context, compressed sensing (CS) emerges as a superior alternative~\cite{wang2023sparse}; by identifying a sparse set of virtual signal sources, CS-based modeling aligns with the underlying physics of wireless propagation while maintaining low computational complexity.
\begin{figure}[!t]
\centerline{\includegraphics[width=\linewidth,trim={1.3cm 7.8cm 2.0cm 8.1cm},clip]{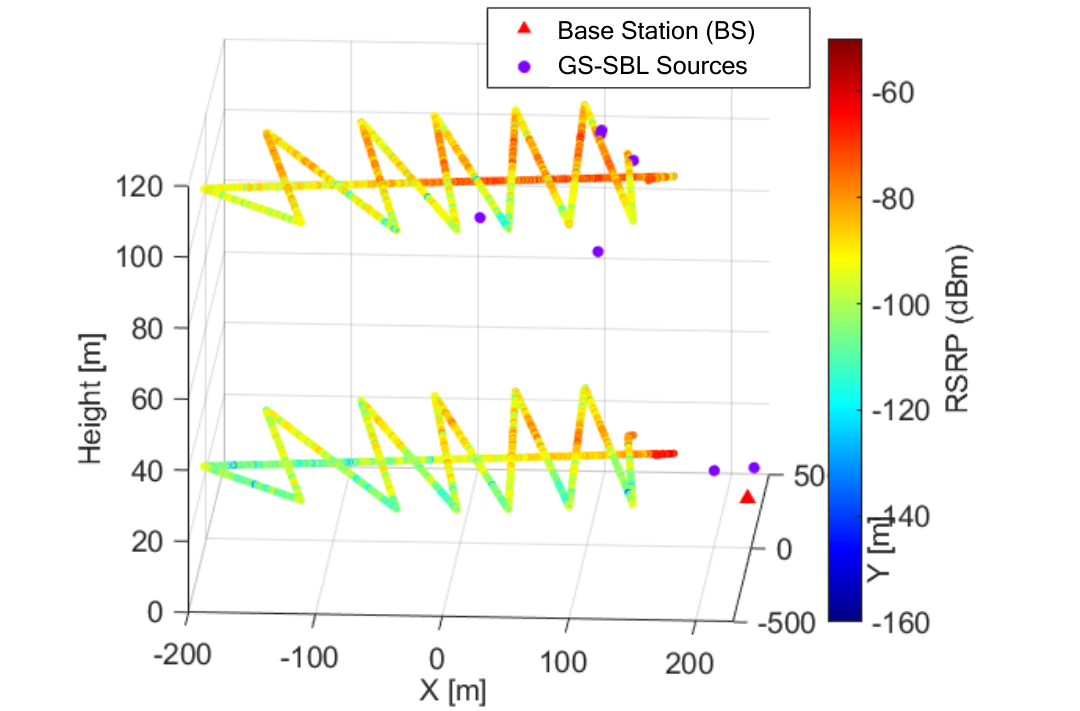}}
\caption{UAV-based signal strength measurements at 30~m and 110~m altitudes. The red triangle ($\triangle$) indicates the physical signal source, while the violet circles ($\circ$) denote the six predicted ``virtual'' source centers ($N_{\text{SBL}}=6$), each representing a 25\,m $\times$ 25\,m $\times$ 10\,m spatial cell.}
\label{fig:uav_trajectory}
\vspace{-6mm}
\end{figure}

In CS-based recovery, signal measurements are decomposed into a propagation-governed sensing matrix and a sparse vector representing potential source powers~\cite{wang2023sparse}. Since the number of actual signal sources is typically negligible compared to the vast 3D spatial grid, the majority of the entries in the sparse vector are expected to be zero. Consequently, the final radio environment is modeled as the superposition of REMs generated by the ``virtual" sources detected by the recovery algorithm. Traditional non-Bayesian methods, such as the Least Absolute Shrinkage and Selection Operator (LASSO)~\cite{bazerque2011group} and Orthogonal Matching Pursuit~(OMP)~\cite{lee2016channel, shen20213d}, have been widely adopted for this purpose. While LASSO provides a global convex optimization framework of the whole sparse vector with $\ell_1$-norm regularization, its computational demands are often prohibitive for real-time applications. Conversely, OMP offers a high-speed greedy alternative, selecting sources sequentially based on the correlation between the sensing features and the signal residual.

Recently, Sparse Bayesian Learning (SBL) has emerged as a preferred framework for CS-based channel modeling~\cite{tipping2003fast, aminu2018bayesian, wang2023sparse, wang2024sparse} due to its superior ability to handle highly correlated sensing matrices—a common characteristic of dense wireless environments. By imposing a hierarchical prior on the source hyperparameters, SBL provides robust uncertainty quantification and better pruning capabilities. However, the computational complexity of standard SBL, typically denoted as $O(M^2 N I)$ (where $M$ is the number of measurements, $N$ is the number of potential sources, and $I$ is the number of iterations), grows rapidly as the search grid expands~\cite{tipping2003fast}. 
Furthermore, while several SBL variants aim to mitigate this computational overhead~\cite{aminu2018bayesian, wang2023sparse, wang2024sparse, zou2025two,pote2023light}, their performance is largely validated on synthetic datasets where pruning is relatively easier.
In real-world scenarios with measurement noise and propagation model mismatch, standard SBL frequently struggles to converge to a sufficiently sparse support set. 

To address these challenges, we propose a novel greedy pursuit strategy that adopts a bottom-up approach to source selection. While conceptually similar to OMP, our method replaces the simple correlation-based selection with a localized ``Micro-SBL'' architecture. For each candidate location, we execute a single-source SBL loop to iteratively calculate the posterior power distribution. The source that achieves the minimum $L_2$ residual error is selected and added to the support set. This process is repeated for $N_{\text{SBL}}$ iterations on the signal residual, where $N_{\text{SBL}}$ denotes the desired number of virtual sources (sparsity level). This hybrid framework maintains the speed of greedy pursuits while leveraging the robust estimation power of Bayesian inference. The primary contributions of this work are summarized as follows: 
\begin{itemize}
    \item We introduce \textbf{GS-SBL} (Greedy Sequential SBL), a novel and efficient framework for CS recovery. This framework reduces SBL complexity by decoupling the simultaneous hyperparameter estimation for \(N\) potential sources into \(N\) single-source Bayesian optimization problems. 
    \item We propose a simple yet effective refinement strategy for the posterior mean powers of the \(N_{\text{SBL}}\) sources, as provided by the ``Micro-SBL" architectures. This strategy avoids solving least squares for the \(N_{\text{SBL}}\) sources.
    \item We outline a computationally efficient Bayesian-powered GS-SBL, less complex or comparable to the non-Bayesian greedy OMP. While correlation-based selection in OMP is rapid, its requirement for repeated least-squares solutions for power estimation becomes computationally extensive as the support set grows.
    \item We show that GS-SBL outperforms OMP and generalizes better to test datasets, particularly when the test datasets are offset by 20~m from the training measurements, using experimental 3D propagation data.
\end{itemize}

\section{System Preliminaries}

\subsection{3D Radio Propagation Model}
We consider a 3D spatial region of interest discretized into a grid of $N$ voxels, where $N = N_x \times N_y \times N_z$. Each voxel represents a potential virtual signal source location. Let $\mathbf{y} \in \mathbb{R}^{M \times 1}$ denote the vector of received signal strength (RSS) measurements in Watts, collected at $M$ distinct sampling locations, where typically $M \ll N$. The relationship between the observed signals and the source powers $\mathbf{x} \in \mathbb{R}^{N \times 1}$ is modeled as a linear system as follows:
\begin{equation}
    \mathbf{y} = \mathbf{\Phi}\mathbf{x} + \mathbf{\epsilon},
\end{equation}
where $\mathbf{\Phi} \in \mathbb{R}^{M \times N}$ is the sensing matrix. The $(i,j)$-th element of $\mathbf{\Phi}$, representing the path loss from the $j$-th virtual source to the $i$-th measurement point, is governed by the Free-Space Path Loss (FSPL) model as follows:
\begin{equation}
    \Phi_{i,j} = G_\text{t}G_\text{r}\left( \frac{\lambda}{4\pi d_{i,j}} \right)^2,
\end{equation}
where $G_\text{t}$ and $G_\text{r}$ are antenna corresponding gains, $\lambda$ is the wavelength, and $d_{i,j}$ is the 3D Euclidean distance between the $i$-th sensor and $j$-th voxel. The term $\mathbf{\epsilon}$ represents additive white Gaussian noise (AWGN) following a normal distribution $p(\mathbf{\epsilon}) \sim \mathcal{N}(0, \sigma_0^2 \mathbf{I})$, where the noise variance is defined by the precision parameter $\beta$ as $\sigma_0^2 = \beta^{-1}$.

\subsection{Sparse Bayesian Learning (SBL)}
In the SBL framework, sparsity is enforced by assuming a hierarchical prior on the source power vector $\mathbf{x}$. We assume each coefficient $x_i$ follows a zero-mean Gaussian distribution:
\begin{equation}
    p(x_i | \alpha_i) = \mathcal{N}(x_i | 0, \alpha_i^{-1}),
\end{equation}
where $\boldsymbol{\alpha} = [\alpha_1, \dots, \alpha_N]^T$ is a vector of precision hyperparameters. To further promote sparsity, a Gamma distribution prior is imposed on $\boldsymbol{\alpha}$ given by:
\begin{equation}
    p(\alpha_i; a, b) = \Gamma(\alpha_i | a, b) = \frac{b^a}{\Gamma(a)} \alpha_i^{a-1} \exp(-b\alpha_i),
\end{equation}
where $a$ and $b$ are small hyperparameters (typically $a, b \to 0$). Given the observed data $\mathbf{y}$ and fixed hyperparameters, the posterior distribution of the sources $p(\mathbf{x} | \mathbf{y}, \boldsymbol{\alpha}, \beta)$ is also Gaussian, $\mathcal{N}(\mathbf{x} | \boldsymbol{\mu}, \mathbf{\Sigma})$, with the mean and covariance defined as:
\begin{equation}
    \mathbf{\Sigma} = (\beta \mathbf{\Phi}^T \mathbf{\Phi} + \mathbf{A})^{-1}, \quad 
    \mathbf{\mu} = \beta \mathbf{\Sigma} \mathbf{\Phi}^T \mathbf{y}, \label{eq:mu_calc}
\end{equation}
where $\mathbf{A} = \text{diag}(\alpha_1, \dots, \alpha_N)$. The recovery process then proceeds by iteratively updating the hyperparameters $\boldsymbol{\alpha}$ and $\beta$, which in turn refine the posterior estimates $\boldsymbol{\mu}$ and $\boldsymbol{\Sigma}$~\cite{wang2024sparse}. This cycle repeats until convergence or until a maximum number of iterations $I$ is reached.

\section{Proposed GS-SBL Methodology}
\subsection{Problem Formulation}
The objective is to identify a sparse support set of size $N_{\text{SBL}}$, representing the most significant virtual signal sources in the 3D environment. Let $\mathcal{I} = \{i_1, i_2, \dots, i_{N_{\text{SBL}}}\}$ denote the indices of the selected sources where $i_k \in \{1, \dots, N\}$, and let $\mathbf{p} = [p_1, p_2, \dots, p_{N_{\text{SBL}}}]^T$ be their corresponding estimated powers. The reconstructed signal at the measurement locations, $\mathbf{\hat{y}}$, is modeled as the linear superposition of the contributions from these selected sources as follows:
\begin{equation}
    \mathbf{\hat{y}} = \sum_{j=1}^{N_{\text{SBL}}} \boldsymbol{\phi}_{i_j} p_j,
\end{equation}
where $\boldsymbol{\phi}_{i_j}$ denotes the $i_j$-th column of the sensing matrix $\mathbf{\Phi}$. The reconstruction task is formulated as a joint optimization problem to minimize the $L_2$ residual error given by:
\begin{equation}
    \{\mathcal{I}^*, \mathbf{p}^*\} = \arg\min_{\mathcal{I}, \mathbf{p}} \| \mathbf{y} - \sum_{j=1}^{N_{\text{SBL}}} \boldsymbol{\phi}_{i_j} p_j \|_2^2.
\end{equation}

\subsection{Greedy Sequential Source Identification}
To solve the optimization problem efficiently, we propose a greedy pursuit strategy that identifies sources sequentially. For the $k$-th source ($k = 1, \dots, N_{\text{SBL}}$), we first compute the residual signal $\mathbf{y}_{\text{res}}^{(k)}$ from the previous stage as follows:
\begin{equation}
    \mathbf{y}_{\text{res}}^{(k)} = \max\left(0,\,\mathbf{y} - \sum_{j=1}^{k-1} \boldsymbol{\phi}_{i_j} p_j\right),
\end{equation}
where the $\max(0, \cdot)$ operator ensures that the residual remains non-negative. This aligns with the physical constraint that source powers and their resulting contributions to the received signal strength must be positive, preventing the algorithm from being misled by negative numerical artifacts in the residual. For the initial step ($k=1$), the residual is simply $\mathbf{y}_{\text{res}}^{(1)} = \mathbf{y}$.

For each candidate source $j \in \{1, \dots, N\}$, we initialize a ``Micro-SBL'' architecture using the $j$-th column $\boldsymbol{\phi}_j$. The hyperparameters are initialized with typical non-sensitive values: $\alpha_j = 0$, $\beta_j = 10^3$, and $a=b=0.05$. Because the problem is reduced to a single-source optimization, the SBL loop converges rapidly; we employ a fixed iteration limit $I = 10$. After the Micro-SBL converges for all $N$ candidates independently, we obtain the posterior mean $\mu_j$ for each candidate using~\eqref{eq:mu_calc} and calculate the $L_2$ residual error as follows:
\begin{equation}
    \mathcal{E}_j = \| \mathbf{y}_{\text{res}}^{(k)} - \mu_j \boldsymbol{\phi}_j \|_2^2.
\end{equation}
The $k$-th source index $i_k$ and its preliminary power $p_k$ are then selected as follows:
\begin{equation}
    i_k = \arg\min_{j} \mathcal{E}_j, \quad p_k = \mu_{i_k}.
\end{equation}

\subsection{Power Refinement and Bias Compensation}
The sequential identification process ensures that each added source reduces the global residual error. However, because the powers are estimated independently in each Micro-SBL loop rather than simultaneously, the vector $\mathbf{p}$ may contain a cumulative estimation bias. To maintain computational efficiency and avoid the complexity of a full Least Squares (LS) optimization over $N_{\text{SBL}}$ sources, we propose a scalar refinement strategy. We define the unbiased power vector as $\mathbf{p}_{\text{unbiased}} = \rho \mathbf{p}$, where $\rho \in (0, 1]$ is a scaling factor that preserves the power ratios between the identified sources. The optimal scalar $\rho^*$ is determined by solving a one-dimensional LS problem as follows:
\begin{equation}
    \rho^* = \arg\min_{\rho} \| \mathbf{y} - \rho \sum_{j=1}^{N_{\text{SBL}}} \boldsymbol{\phi}_{i_j} p_j \|_2^2
\end{equation} 
This refinement ensures the global signal level is correctly calibrated while significantly reducing the total computational footprint compared to standard SBL or multi-variable LS.

\section{Experimental Results and Analysis}
\subsection{3D wireless Propagation Dataset}
\label{sec:datasets}
The GS-SBL framework is evaluated against a publicly available empirical UAV-based reference signal received power~(RSRP) dataset~\cite{IEEEDataPort_2}. The experimental setup for this dataset involved a fixed base station (BS) with a 10\,m antenna height serving as the transmitter, which operated at a center frequency of 3.5\,GHz and a 1.4\,MHz bandwidth. Measurements were recorded via a UAV following zigzag flight patterns at five distinct altitudes, starting at 30\,m and increasing by 20\,m up to 110\,m. The available RSRP distributions at the 30\,m and 110\,m altitudes are illustrated in Fig.~\ref{fig:uav_trajectory}.


\subsection{Performance Analysis and Source Localization}
\begin{figure}[!t]
\centerline{\includegraphics[width=\linewidth,trim={1.3cm 6.4cm 2.0cm 7.1cm},clip]{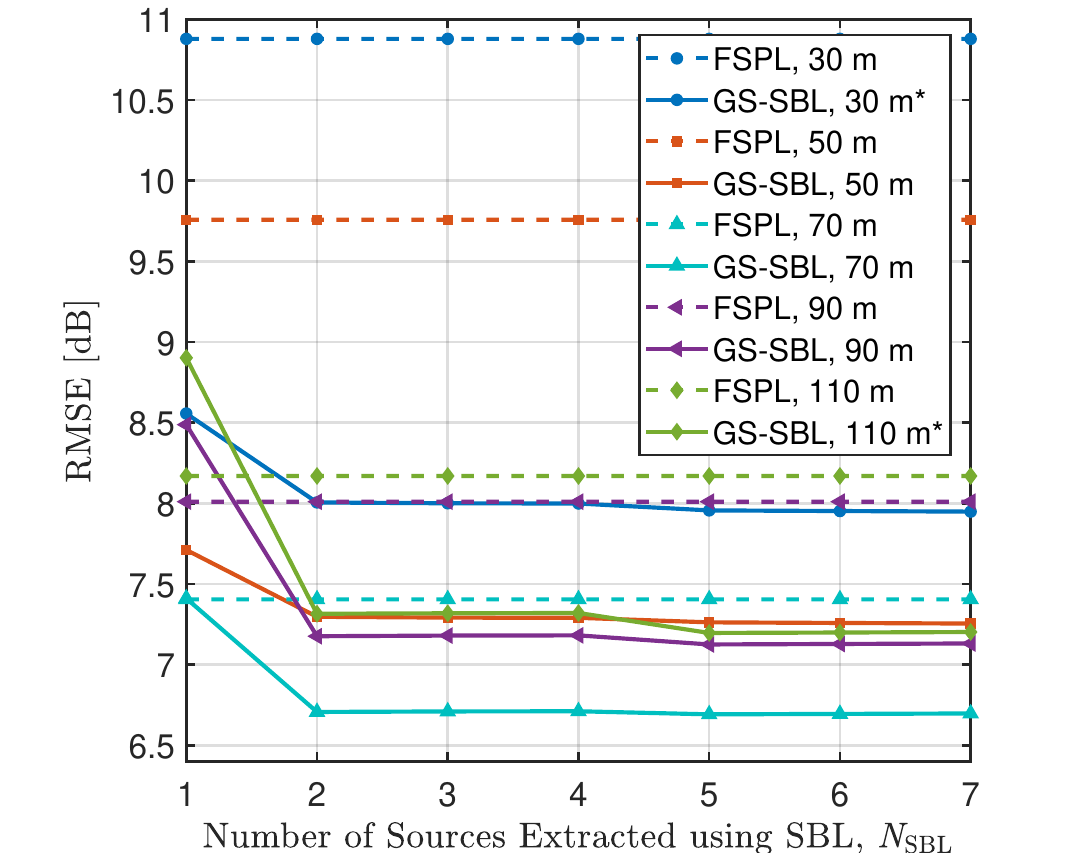}}
\caption{Impact of the number of virtual sources $N_{\text{SBL}}$ on GS-SBL performance, measured in RMSE, across five measurement altitudes. Training altitudes (30\,m and 110\,m) are indicated by an asterisk (*) in the legend.}
\label{fig:rmse_saturation}
\vspace{-4mm}
\end{figure}

To determine the optimal number of virtual sources required to characterize the 3D environment, we evaluate the reconstruction performance for $N_{\text{SBL}}$ values ranging from 1 to 7. As illustrated in Fig.~\ref{fig:rmse_saturation}, the root mean square error~(RMSE) significantly decreases as $N_{\text{SBL}}$ increases from 1 to 2, after which the error reduction saturates.
This suggests that two virtual sources are sufficient to capture the dominant propagation characteristics of this scenario. While the GS-SBL framework was trained using measurements from 30\,m and 110\,m altitudes, it consistently outperforms the FSPL baseline across all test altitudes (50\,m, 70\,m, and 90\,m), demonstrating robust predictive capabilities at unsampled elevations.

\begin{figure}[t!]
    \centering
    \begin{subfigure}{0.23\textwidth}
        \centering
        \includegraphics[width=\linewidth,trim={2.5cm 7.0cm 3.5cm 8.0cm},clip]{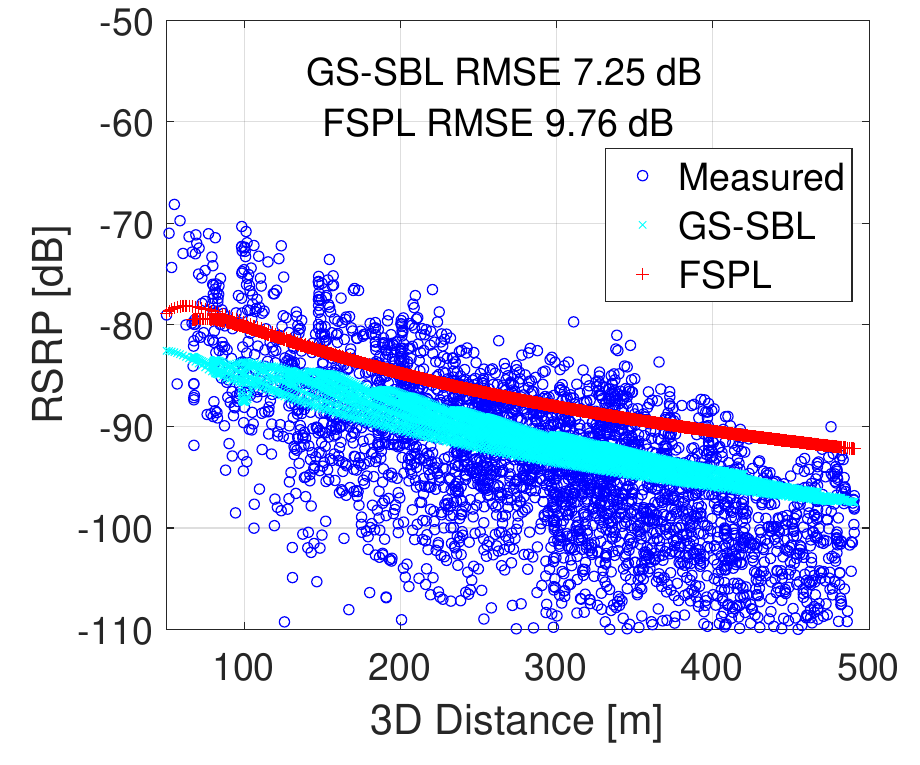}
        \caption{50~m}
    \end{subfigure}
    \begin{subfigure}{0.23\textwidth}
        \centering
        \includegraphics[width=\linewidth,trim={2.5cm 7.0cm 3.5cm 8.0cm},clip]{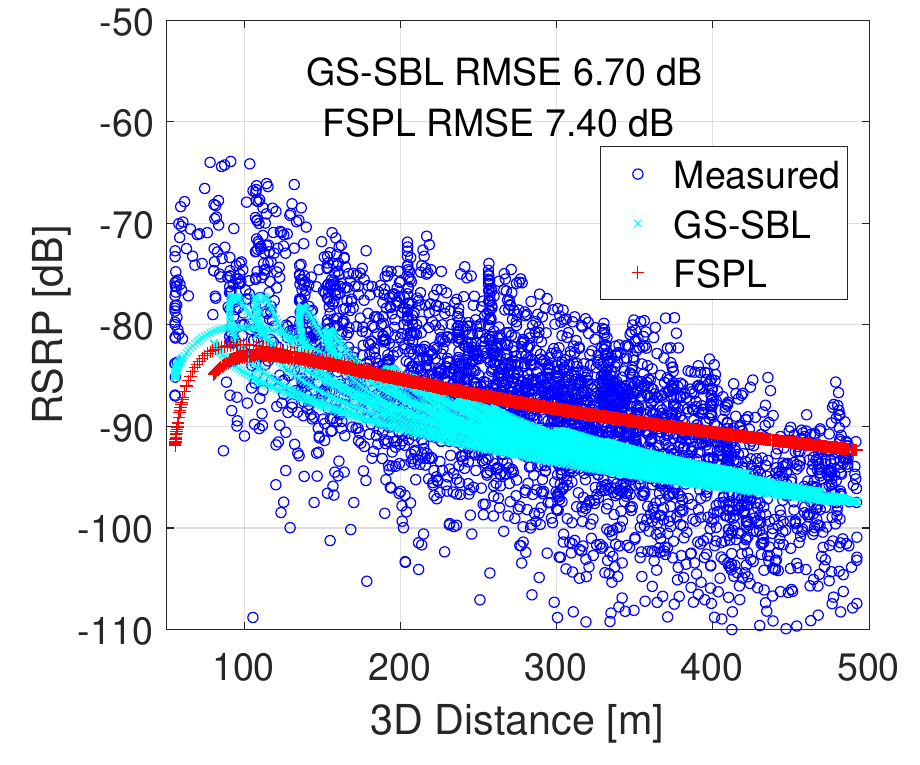}
        \caption{70~m}
    \end{subfigure}
    \begin{subfigure}{0.23\textwidth}
        \centering
        \includegraphics[width=\linewidth,trim={2.5cm 7.0cm 3.5cm 8.0cm},clip]{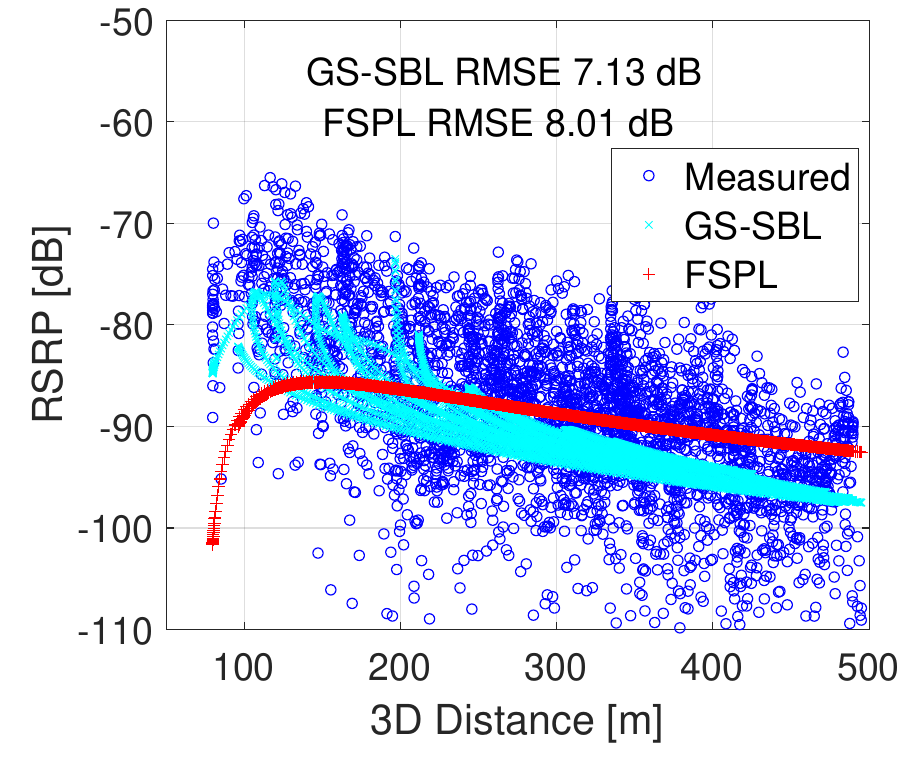}
        \caption{90~m}
    \end{subfigure}
    \begin{subfigure}{0.23\textwidth}
        \centering
        \includegraphics[width=\linewidth,trim={2.5cm 7.0cm 3.5cm 8.0cm},clip]{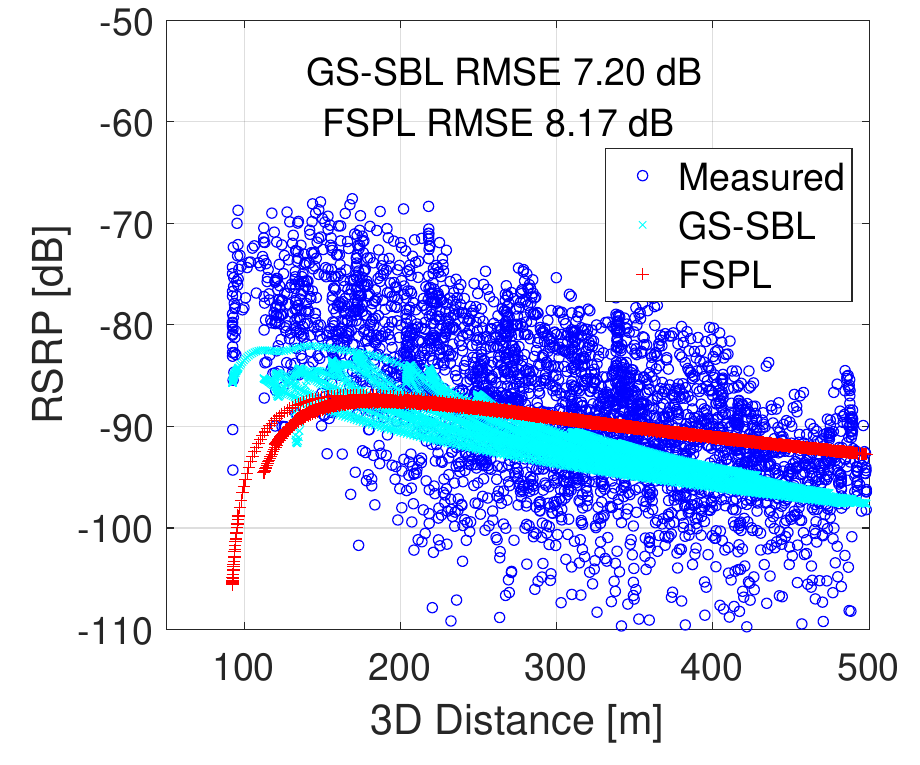}
        \caption{110~m*}
    \end{subfigure}
    \vspace{-1mm}
        \caption{Measured RSRP versus 3D distance compared with FSPL and the GS-SBL predictions at four altitudes. The training altitude for GS-SBL is marked with an asterisk (*).}
    \label{fig:scatter_3d_plots}\vspace{-4mm}
\end{figure}

To interpret these results physically, we analyze the relationship between 3D distance and RSRP, as shown in Fig.~\ref{fig:scatter_3d_plots}.
By examining RSRP peaks that deviate from the average FSPL trend—most notably at the 90\,m altitude—we can infer the presence of the secondary virtual source. The GS-SBL framework identifies this source's spatial coordinates by minimizing the global $L_2$ residual, placing it at a location that accounts for the observed constructive interference and high-signal variance in the empirical data at higher altitudes.

\begin{figure}[!t]
\centerline{\includegraphics[width=0.95\linewidth,trim={3.55cm 10.9cm 4.25cm 11.0cm},clip]{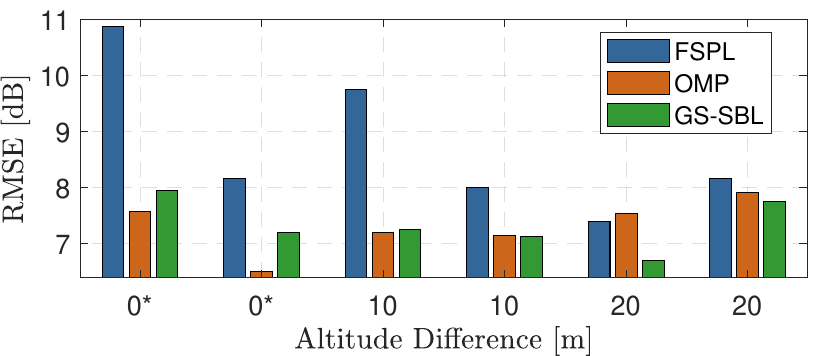}}
\caption{RMSE of the proposed GS-SBL, OMP, and FSPL under training-test elevation separations of $\{0, 10, 20\}$\,m. An asterisk (*) on the X-axis indicates performance on the training set (0\,m), while the rest represent the model's generalization performance on unseen test altitudes.}
\label{fig:comparison_omp}
\vspace{-6mm}
\end{figure}

Finally, we assess the effectiveness of GS-SBL relative to existing CS techniques. While LASSO provides a convex optimization alternative, its high computational complexity makes it impractical for real-time 3D REM construction. Consequently, we compare our framework against OMP, a standard non-Bayesian greedy pursuit algorithm. Fig.~\ref{fig:comparison_omp} illustrates the RMSE for both techniques as a function of the elevation separation between the training and test sets.
At 0\,m separation, OMP exhibits slightly lower error compared to GS-SBL, likely due to overfitting the training noise. However, as the elevation separation increases to 10\,m and 20\,m, the performance of OMP degrades significantly. In contrast, GS-SBL maintains a lower and more stable RMSE, indicating that the Bayesian-powered approach achieves superior generalization by learning the underlying propagation physics rather than localized measurement fluctuations.

\section{Conclusion}
In this work, we presented GS-SBL, a novel Greedy Sequential Sparse Bayesian Learning framework designed for efficient 3D wireless path loss modeling. By introducing a ``Micro-SBL'' architecture, we successfully decoupled the high-dimensional hyperparameter estimation of standard SBL into multiple single-source optimization problems. This approach bridges the gap between the computational speed of greedy pursuits and the robust uncertainty quantification of Bayesian inference. Validated against empirical 3D RSRP datasets from a UAV-based setup, GS-SBL demonstrated a significant performance enhancement. Crucially, our results reveal that while non-Bayesian methods like OMP tend to overfit localized training noise, GS-SBL identifies physically consistent virtual sources, leading to superior generalization capabilities across different altitudes. 
Future research will focus on expanding the comparative analysis to include other advanced SBL variants, such as Clustered or Block-Sparse SBL, to evaluate performance in even more densely cluttered environments. Additionally, we intend to investigate horizontal spatial separation between training and testing sets to further stress-test the model's geographical extrapolation.

\section*{ACKNOWLEDGEMENT}
We use measurement data published by the NSF Aerial Experimentation and Research Platform on Advanced Wireless (AERPAW) platform. This research is supported in part by the NSF award CNS-2332835 and the INL Laboratory Directed Research Development (LDRD) Program under BMC No. 264247, Release No. 26 on BEA’s Prime Contract No. DE-AC07-05ID14517.

\bibliographystyle{IEEEtran}
\bibliography{ref}

\end{document}